\documentstyle[twocolumn,aps,epsfig]{revtex}
\begin{document}
\draft
\preprint{\vbox{\hbox{IFT--P.085/98}\hbox{hep-ph/9811373}}}
\twocolumn[\hsize\textwidth\columnwidth\hsize\csname @twocolumnfalse\endcsname
\title{New Higgs Couplings at \boldmath{$e^+ e^-$} and Hadronic Colliders}
\author{M.\ C.\ Gonzalez--Garcia $^{1,2}$, 
	S.\ M.\ Lietti $^2$ and 
	S.\ F.\ Novaes $^2$}
\address{
$^1$ Instituto de F\'{\i}sica Corpuscular IFIC/CSIC,
     Departament de F\'{\i}sica Te\`orica\\
     Universitat de Val\`encia, 46100 Burjassot, Val\`encia, Spain\\
$^2$ Instituto de F\'{\i}sica Te\'orica, 
     Universidade  Estadual Paulista, \\  
     Rua Pamplona 145, 01405--900, S\~ao Paulo, Brazil.}
\date{\today}
\maketitle
\widetext
\begin{abstract}
We examine the potentiality of both CERN LEP and Fermilab Tevatron
colliders to establish bounds on new couplings involving the
bosonic sector of the standard model. We pay particular attention
to the anomalous Higgs interactions with $\gamma$, $W^{\pm}$ and
$Z^0$. A combined exclusion plot for the coefficients of different
anomalous operators is presented. The sensitivity that can be
achieved at the Next Linear Collider and at the upgraded Tevatron
is briefly discussed. 
\end{abstract}

\vskip2pc]
\narrowtext
\section{Introduction: Effective Lagrangians for Higgs Interactions} 
\label{int}

We certainly expect that the standard model (SM), despite its
astonishing success in describing all the precision high energy
experimental data \cite{ichep98}, to be an incomplete picture of
Nature at high energy scales. In particular, the Higgs sector of
the model, responsible for the spontaneous electroweak symmetry
breaking and for the mass generation, is introduced in an {\it ad
hoc} way and has not yet been directly probed.

Although we do not know the specific theory which will eventually
supersede the SM, we can always parametrize its effects by means
of an effective Lagrangian \cite{effective} that contains
operators with dimension higher than four and involves the fields
and symmetries of the low energy theory. The effective Lagrangian
approach is a model--independent way to describe new physics that
is expected to manifest itself directly at an energy scale
$\Lambda$,  larger than the scale where the experiments are
performed.

The effective Lagrangian depends on the particle content at low
energies. We consider here the possibility of having a light
Higgs boson that should be present in the higher dimensional
operators. Hence, we use a linearly realized $SU_L(2) \times
U_Y(1)$ invariant effective Lagrangian \cite{linear,hisz} to
describe the bosonic sector of the SM, keeping the fermionic
sector unchanged. 

A general set of dimension--6 operators that involve  gauge
bosons and the Higgs scalar field, respecting local $SU_L(2)
\times U_Y(1)$ symmetry, and $C$ and $P$ conserving, contains
eleven  operators \cite{linear}. Some of these operators either
affect only the Higgs self--interactions or contribute to the
gauge boson two--point functions at tree level and is 
strongly constrained from low energy physics below the present
sensitivity of high energy experiments \cite{hisz}. The remaining
five ``blind'' operators can be written as
\cite{linear,hisz},
\begin{eqnarray}
&& {\cal L}_{\text{eff}} = \sum_i \frac{f_i}{\Lambda^2} {\cal O}_i 
= \frac{1}{\Lambda^2} \Bigl[ 
f_{WWW}\, Tr[\hat{W}_{\mu \nu}\hat{W}^{\nu\rho}\hat{W}_{\rho}^{\mu}] 
\nonumber \\
&&+ f_W (D_{\mu} \Phi)^{\dagger} \hat{W}^{\mu \nu} (D_{\nu} \Phi) 
+ f_B (D_{\mu} \Phi)^{\dagger} \hat{B}^{\mu \nu} (D_{\nu} \Phi) 
\nonumber \\ 
&&+ f_{WW} \Phi^{\dagger} \hat{W}_{\mu \nu} \hat{W}^{\mu \nu} \Phi  
+ f_{BB} \Phi^{\dagger} \hat{B}_{\mu \nu} \hat{B}^{\mu \nu} \Phi 
  \Bigr] 
\label{lagrangian}
\end{eqnarray}
where $\Phi$ is the Higgs field doublet,  
$\hat{B}_{\mu\nu} = i (g'/2) B_{\mu \nu}$, and $\hat{W}_{\mu \nu} = i (g/2)
\sigma^a W^a_{\mu \nu}$
with $B_{\mu \nu}$ and $ W^a_{\mu \nu}$ being the field strength
tensors of the $U(1)$ and $SU(2)$ gauge fields respectively. 

Anomalous $H\gamma\gamma$, $HZ\gamma$, and $HZZ$ and $HWW$ and 
couplings are generated by (\ref{lagrangian}), which modify the 
Higgs boson production and
decay \cite{hagiwara2}. In the unitary gauge they are given by 
\begin{eqnarray}
{\cal L}_{\text{eff}}^{\text{H}} &=& 
g_{H \gamma \gamma} H A_{\mu \nu} A^{\mu \nu} + 
g^{(1)}_{H Z \gamma} A_{\mu \nu} Z^{\mu} \partial^{\nu} H \nonumber \\ 
&+& g^{(2)}_{H Z \gamma} H A_{\mu \nu} Z^{\mu \nu}
+ g^{(1)}_{H Z Z} Z_{\mu \nu} Z^{\mu} \partial^{\nu} H \nonumber \\
&+& g^{(2)}_{H Z Z} H Z_{\mu \nu} Z^{\mu \nu} +
g^{(2)}_{H W W} H W^+_{\mu \nu} W_{-}^{\mu \nu} \; \nonumber \\
&+&g^{(1)}_{H W W} \left (W^+_{\mu \nu} W_{-}^{\mu} \partial^{\nu} H 
+h.c.\right)\,
\label{H} 
\end{eqnarray}
where $A(Z)_{\mu \nu} = \partial_\mu A(Z)_\nu - \partial_\nu
A(Z)_\mu$. The effective couplings $g_{H \gamma \gamma}$,
$g^{(1,2)}_{H Z \gamma}$, and $g^{(1,2)}_{H Z Z}$  and 
$g^{(1,2)}_{H WW}$ are related
to the coefficients of the operators appearing in (1) and 
can be found 
elsewhere \cite{hagiwara2}. In particular the Higgs couplings
to two photons is given by 
\begin{eqnarray}
g_{H \gamma \gamma} &=& - \left( \frac{g M_W}{\Lambda^2} \right)
                       \frac{s^2 (f_{BB} + f_{WW})}{2} \; 
\label{g} 
\end{eqnarray}
with $g$ being the electroweak coupling constant, and $s(c)
\equiv \sin(\cos)\theta_W$. 

Equation (\ref{lagrangian}) also generates  new contributions to
the  triple gauge boson vertex \cite{linear,hisz}. The operators
${\cal O}_{W}$  and ${\cal O}_{B}$ give rise to both anomalous
Higgs--gauge boson couplings and to new triple and quartic
self--couplings amongst the gauge bosons, while the operator
${\cal O}_{WWW}$ solely modifies the gauge boson
self--interactions. On the other hand ${\cal O}_{WW}$ and ${\cal
O}_{BB}$ only affect $HVV$ couplings, since their contribution to
the $WW\gamma$ and $WWZ$ tree--point couplings can be completely
absorbed in the redefinition of the SM fields and gauge couplings
\cite{hisz,hagiwara2}.  Therefore, one cannot obtain any
constraint on these couplings from the  study of anomalous
trilinear gauge boson couplings.  Finally, we should point out that
the dimension-six operators (\ref{lagrangian})  do not induce
$4$--point anomalous couplings like $Z Z \gamma \gamma$,  $Z
\gamma\gamma \gamma$, and $\gamma \gamma \gamma \gamma$, being
these  terms generated only by dimension--eight and higher
operators. 

Anomalous Higgs boson couplings have been studied in Higgs and
$Z^0$ boson decays \cite{hagiwara2}, and in $e^+ e^-$
\cite{ee,our,our:lep2aaa,our:NLCWWA,our:NLCZZA}, $p \bar{p}$
\cite{our:tevatronjj,our:tevatronmis,our:tevatron3a} and
$\gamma\gamma$ colliders \cite{gamma}. In this work, we make a
combined analysis, based on several experimental searches at the
CERN LEP collider and at the Fermilab Tevatron collider, in order
to establish the attainable bounds on the coefficient of the
effective operators describing the anomalous bosonic sector.  Our
results are presented in Section \ref{today}. In Section
\ref{future} we discuss the sensitivity that can be achieved at
the Fermilab Tevatron upgrade and at the Next Linear Collider
(NLC). Finally, in Section \ref{Conclusions},  we compare our
results with existing limits on the  coefficients of
dimension-six operators based on searches for anomalous  triple
gauge boson couplings.


\section{Bounds from the Recent LEP and Tevatron Searches}
\label{today}

In this section, we derive  combined bounds on anomalous Higgs
boson interactions taking into account both LEP \cite{opal:ggg}
and Tevatron \cite{d0jj,d0miss,cdf} data on the following
signatures:
\begin{eqnarray}
e^+\, e^-  & \rightarrow & \gamma\, \gamma\, \gamma \nonumber \\
p\, \bar p & \rightarrow & j \, j \, \gamma\, \gamma  \label{proc}
\\
p\, \bar p & \rightarrow & \gamma \,\gamma \,+ \,\not \!\! E_T \nonumber \\
p\, \bar p & \rightarrow & \gamma\, \gamma\, \gamma \nonumber
\end{eqnarray} 

Events containing two photons plus missing energy, additional
photons or charged fermions represent a signature for several
theories involving physics beyond the SM, such as some classes of
supersymmetric models \cite{xer} and they have been extensively
searched for \cite{opal:ggg,d0jj,d0miss,cdf}. In the framework of
anomalous Higgs couplings presented before, they can also arise
from the production of a Higgs boson which subsequently decays in
two photons.  In the SM, the  decay width $H \to \gamma \gamma$
is very small since it occurs just at one--loop level
\cite{h:gg}. However, the existence of the new interactions
(\ref{H}) can enhance this width in a significant way. Recent
analyses of these signatures presented a good agreement with  the
expectations from the SM. Thus we can employ these negative
experimental  results  to constrain new anomalous couplings in
the bosonic sector of the SM. 

We have included in our calculations all SM (QCD plus
electroweak), and anomalous contributions that lead to these
final states. The SM one-loop contributions to the
$H\gamma\gamma$ and $H Z\gamma$ vertices were introduced through
the use of the  effective operators with the corresponding form
factors in the coupling.  Neither the narrow--width approximation
for the Higgs boson  contributions, nor the effective $W$ boson
approximation were employed. We consistently included the effect
of all interferences between the anomalous signature and the SM
background.  The SM Feynman diagrams corresponding to the
background subprocess were generated by Madgraph \cite{mad} in
the framework of Helas \cite{helas}. The anomalous couplings
arising from the Lagrangian (\ref{lagrangian}) were implemented
in Fortran routines and were included accordingly. For the
$p\,\bar p$ processes, we have used the  MRS (G) \cite{mrs} set
of proton structure functions with the scale $Q^2 = \hat{s}$.

All processes listed in (\ref{proc}) have been the object of
direct experimental searches. In our analysis we have closely
followed theses searches in order to make our study as realistic
as  possible. In order to establish bounds on the values of the
anomalous  coefficients $f_i$, $i=WW,BB,W,B$, we have imposed an
upper limit on  the number of signal (anomalous) events based on
Poisson statistics. In the absence of background this implies
$N_{\text{signal}} < 1 \, (3)$ at 64\% (95\%) CL. In the presence
of background events, we employed the  modified Poisson analysis
\cite{otaviano}.

For events containing three photons in the final  state at
electron--positron collisions \cite{our:lep2aaa},
\begin{eqnarray}
e^+ e^- &\rightarrow& \gamma +H(\rightarrow \gamma \gamma) \; ,
\label{ggg}
\end{eqnarray}
we have used the recent OPAL data \cite{opal:ggg} where data
taken at several energy points in the range $\sqrt{s}=130$ --
$172$ GeV, were  combined. They have established an upper limit
at 95\% CL for $\sigma(e^+ e^- \rightarrow \gamma + X) \times
BR(X \to \gamma \gamma)$ where $X$ is a scalar particle. 
These results were used to derive our limits.

The process
\begin{eqnarray}
p \bar{p} &\to & W (Z) (\to j\, j) + H (\to \gamma \gamma).
\label{jjreaction}  
\end{eqnarray}
can also be employed to further constrain the anomalous Higgs
boson couplings described in (\ref{H}) \cite{our:tevatronjj}. D\O
~Collaboration reported the results for the search of high
invariant--mass photon pairs in $p \bar{p} \to \gamma \gamma j j$
events \cite{d0jj} at $\sqrt{s}=1.8$ TeV and $100$ pb$^{-1}$ of
integrated luminosity. In our analysis, we applied the same cuts
of Ref.\ \cite{d0jj} and included the particle identification and
trigger efficiencies. We have searched for Higgs boson with mass
in the range $100 < M_H \lesssim 220$, since after the $WW(ZZ)$
threshold is reached the diphoton branching ratio of Higgs is
quite reduced. Since no event with two--photon invariant mass in
the range $100 < M_{\gamma\gamma} \lesssim 220$ were observed, a
$95\%$ CL in the determination of the anomalous coefficient $f_i$
is attained requiring 3 events coming only from the anomalous
contributions. 

For events containing two photons plus large missing transverse
energy ($\gamma\gamma \not \!\! E_T$) \cite{our:tevatronmis}  we
have used the results from D\O ~collaborations \cite{d0miss}.
Anomalous Higgs couplings can give rise to this final state via, 
\begin{eqnarray}
p \bar{p} & \to & Z^0 (\to \nu \bar{\nu}) + H (\to \gamma \gamma) + X
\nonumber \\ 
p \bar{p} & \to & W (\to \ell \nu) + H (\to \gamma \gamma) + X
\label{zw}  
\end{eqnarray}
where in the latter case the charged lepton ($\ell = e, \mu$)
escapes undetected. 

\begin{figure}
\protect
\epsfxsize=10cm
\begin{center}
\leavevmode \epsfbox{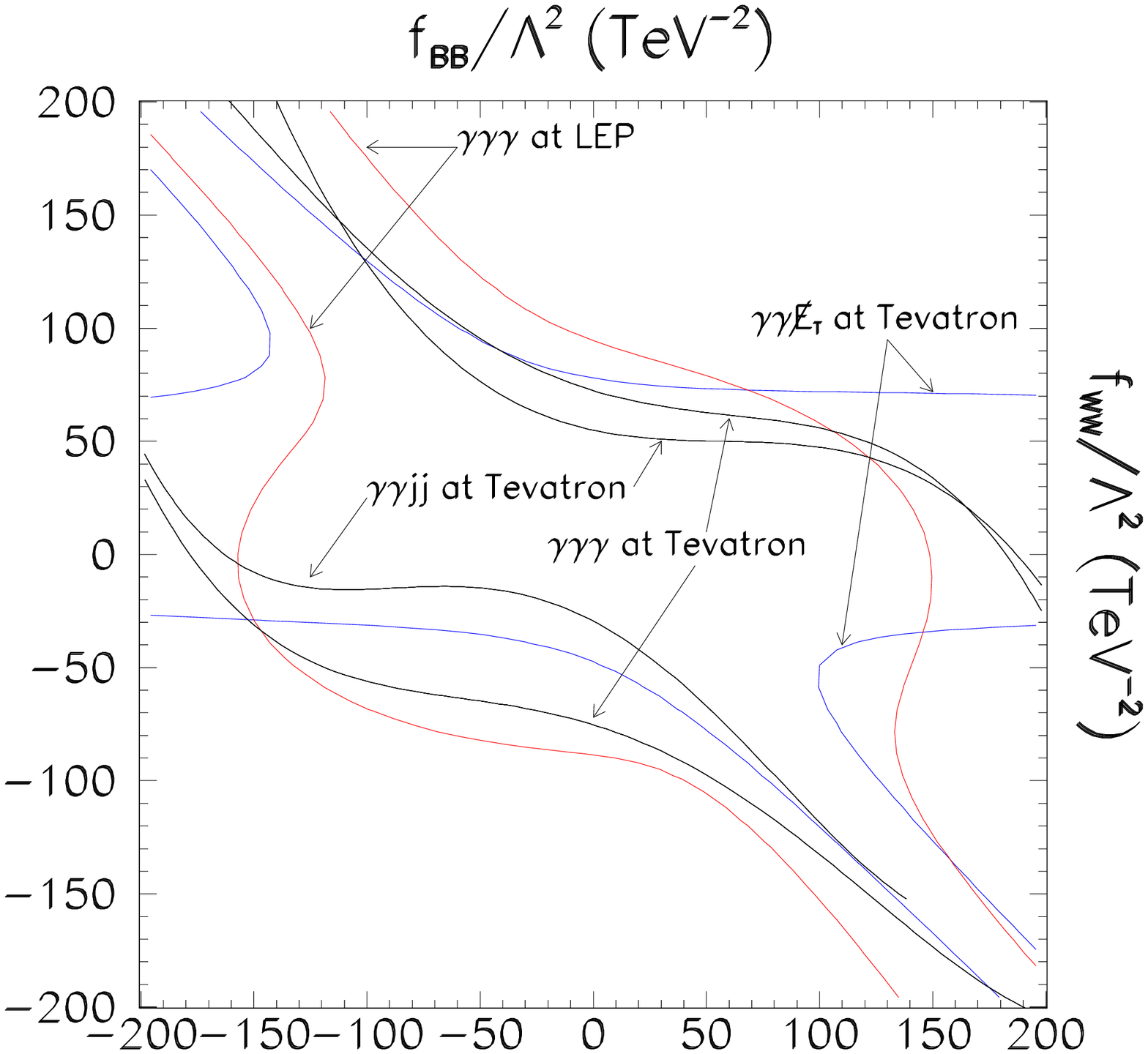}

{\large\bf (a)}
\end{center}
\epsfxsize=10cm
\begin{center}
\leavevmode \epsfbox{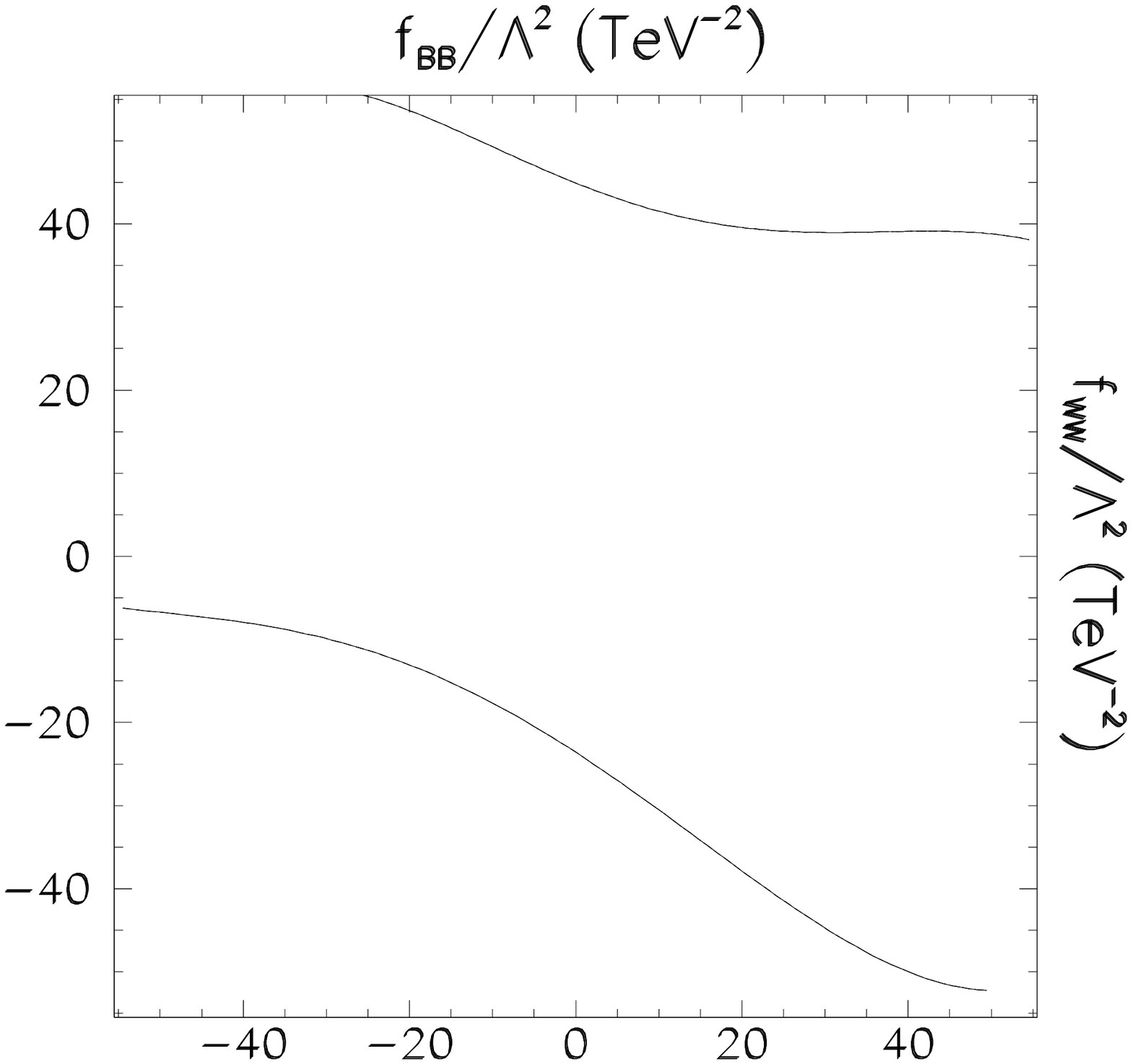}

{\large\bf (b)}
\end{center}
\caption{{\bf (a)} Exclusion region outside the curves in the
$f_{BB} \times f_{WW}$ plane, in TeV$^{-2}$, based on the CDF
analysis \protect\cite{cdf} of $\gamma\gamma\gamma$ production
(most external  black lines), on the D\O ~analysis
\protect\cite{d0jj} of $\gamma\gamma j j$  production (most
internal black lines), on the D\O ~analysis \protect\cite{d0miss}
of $\gamma\gamma \not \!\! E_T$ (blue lines), and on the OPAL
analysis \protect\cite{opal:ggg} of $\gamma\gamma\gamma$
production (red lines) , always assuming $ M_H = 100$ GeV.  The
curves show the 95\% CL deviations from the SM total cross
section.  {\bf (b)} Same as {a} for the combined analysis.}
\label{fig1}
\end{figure}

In order to compare our predictions with the results of  D\O
~Collaboration, we have applied the same cuts of last article in
Ref.\ \cite{d0miss}. After these cuts we find that 80\% to 90\%
of the signal comes from associated Higgs--$Z^0$ production while
10\% to 20\% arrises from Higgs--$W$. We also include in our
analysis the particle identification and trigger efficiencies
which vary from 40\% to 70\% per photon \cite{fermilab}.  Since
no event with two--photon invariant mass in the range $100 <
M_{\gamma\gamma} \lesssim 2 M_W$ were observed, a $95\%$ CL in
the determination of the anomalous coefficient $f_i$, $i=WW, BB,
W, B$ is attained requiring 3 events coming only from the
anomalous contributions. Table \ref{tab:f1} shows  the 95\% CL
allowed region of the anomalous couplings in the above scenario.
We exhibit in Fig.\ \ref{fig1} the 95\% CL exclusion region in
the plane $f_{BB} \times f_{WW}$ obtained from the D\O ~data on
$\gamma \gamma + \not \!\! E_T $ \cite{d0miss}. 

Finally we have also analyzed events with three photons in the
final state \cite{our:tevatron3a}
\begin{eqnarray}
p \bar p &\rightarrow& \gamma +H(\rightarrow \gamma \gamma) \; ,
\label{pp:ggg}
\end{eqnarray}
and compare our results  with the recent search reported by CDF
Collaboration \cite{cdf} for this signature. They looked for
$\gamma\gamma\gamma$ events requiring two photons in the central
region of the detector, with a minimum transverse energy of 12
GeV, plus an additional photon with $E_T > 25$ GeV. The photons
were required to be separated by more than $15^\circ$. In these
conditions they were able to establish that the signal should
have less than 3 events, in the 85 pb$^{-1}$ collected data, at
95 \% CL. 

We have used the results described above to constrain the value
of the coefficients $f_i$ of (\ref{lagrangian}). The coupling
$H\gamma\gamma$ (\ref{g}) involves $f_{WW}$ and $f_{BB}$
\cite{hagiwara2}, and in consequence, the anomalous signature
$f\bar f \gamma\gamma$ is only possible when those couplings are
not vanishing. The couplings $f_B$ and $f_W$, on the other hand,
affect the production mechanisms for the Higgs boson.  In Fig.\
\ref{fig1}(a) we present our results for the excluded  region in
the $f_{WW}$, $f_{BB}$ plane from the different channels  studied
for $M_H=100$ GeV assuming that these are the only non-vanishing
couplings. Since the anomalous contribution to  $H\gamma\gamma$
is zero for $f_{BB} = - f_{WW}$, the bounds become very weak
close to this line, as is clearly shown in Fig.\ \ref{fig1}.

In order to reduce the number of free parameters one can make the
assumption that all blind operators affecting the Higgs
interactions have a common coupling $f$, {\it i.e.} $f = f_W =
f_B = f_{WW} = f_{BB}$ \cite{hisz,hagiwara2,review}. We present
in Table \ref{tab:f1} the 95\% CL allowed regions of the
anomalous couplings in this scenario, for different  Higgs boson
mass. 

These results obtained from the analysis of the four reactions
(\ref{proc}) can be statistically  combined in order to obtain a
better bound on the coefficient of the effective operators
(\ref{lagrangian}).  We exhibit in Fig.\ \ref{fig1}(b) the 95\%
CL exclusion region in the plane $f_{BB} \times f_{WW}$ obtained
from combined results. In Fig.\ \ref{kappa}, we present the
combined limits for the coupling  constant $f = f_{BB}= f_{WW}=
f_{B} = f_{W}$ (upper scale) for Higgs boson masses  in the range
of $100 \leq M_H \leq 220$ GeV. 

\section{Future Perspectives}
\label{future}

The effect of the anomalous operators becomes more evident with
the increase of energy, and higher sensitivity to smaller values of
the anomalous coefficients can be achieved by studying their
contribution to different processes at the upgraded Tevatron
collider or at new machines, like the Next Linear Collider. 

We first extend our analysis of the $p \bar{p} \to \gamma \gamma
\not\!\!E_T$ and $p \bar{p} \to \gamma \gamma j j$ reactions for
the upgraded  Tevatron collider. We have considered the Run II
upgrade with a luminosity of 1 fb$^{-1}$, and for the TeV33
upgrade we  assumed 10 fb$^{-1}$ \cite{tevatron}. In our
estimates we have taken the same cuts and detection efficiencies
given in our previous analysis. 

For the $\gamma\gamma\gamma$ final state we have studied the
improvement on the sensitivity to the anomalous coefficients by
implementing additional kinematical cuts \cite{our:tevatron3a}.
Best results are obtained for the following set of cuts:
$E_{T_{1}} > 40$ GeV, with $E_{T_{2,3}} > 12$ GeV where  we have
ordered the three photons according to their transverse energy,
{\it i.e.\/} $E_{T_{1}} > E_{T_{2}} > E_{T_{3}}$. We always
require the photons to be in the central region of the detector
($|\eta_{i}| < 1$) where there is sensitivity for electromagnetic
showering. In our estimates we assume the same detection
efficiency for photons as considered by CDF Collaboration
\cite{cdf}.

In Table \ref{tab:run2} we present the 95\% CL  limit on the
anomalous couplings for Tevatron Run II and for TeV33  for each
individual process. All couplings are assumed equal  ($f =
f_{BB}= f_{WW}= f_{B} = f_{W}$ ) and the Higgs boson mass is
varied in the range $100 \leq M_H \leq 220$ GeV. Combination of
the results obtained from the analysis of the three  reactions
(\ref{jjreaction},\ref{zw},\ref{pp:ggg}) leads to the improved
bounds given in the last column of Table \ref{tab:run2}.
Comparing these results with those in Table.\ \ref{tab:f1} (or
with the upper scale of  Fig.\ \ref{kappa}) we observe an
improvement of about a factor $\sim 2$--$3$ [$\sim 4$--$6$] for
the combined limits at RunII [TeV33].  

The Next Linear electron--positron Collider will open an
important opportunity to further improve the search for new
physics. In particular, the anomalous Higgs boson couplings can
be investigated in the processes \cite{our:NLCWWA,our:NLCZZA}:
\begin{eqnarray}
e^+ e^- &\to& W^+ W^- \gamma \label{wwg} \\
e^+ e^- &\to& Z^0 Z^0 \gamma  \label{zzg}
\end{eqnarray}

We studied the sensitivity of NLC to these processes assuming a
energy in the center--of--mass of  $\sqrt{s} = 500 \; \text{GeV}$
and an integrated luminosity  ${\cal L} = 50  \;\text{fb}^{-1}$.
We adopted a cut in the photon  energy of $E_\gamma > 20 \;
\text{GeV}$ and required the angle between  any two particles to
be larger than $15^\circ$. We have analyzed these  processes for
different values of the Higgs boson mass.

We have investigated different distributions of the final state
particles in order to search for kinematical cuts that could
improve the NLC sensitivity. The most promising variable is the
photon transverse momentum. We observe that the contribution of
the anomalous couplings is larger in the high $p_{T_\gamma}$
region. Since  the anomalous signal is dominated by
on--mass--shell Higgs$\gamma$ production with the subsequent $H
\to W^+ W^-$ or $Z^0Z^0$ decay, the photon transverse momentum
is distributed around  the monochromatic peak
$E_\gamma^{\text{mono}}= (s - M_H^2)/(2 \sqrt{s})$. In
consequence for  Higgs boson masses in the range $2 M_{W,Z} \leq
M_H \leq (\sqrt{s} - E_\gamma^{\text{min}})$ GeV, where on--shell
production is allowed,   a cut of $p_{T_\gamma} \gtrsim 100$
drastically reduces the background.   For lighter Higgs bosons,
{\it e.g.\/} $M_H <  2 M_{W,Z}$, the $p_{T_\gamma}$ cut is
ineffective since the Higgs boson is off--mass--shell and the
peak in the photon transverse momentum distribution disappears.
This makes the bounds on the anomalous coefficients obtained from
the $W^+ W^- (Z^0Z^0) \gamma$ production to be very loose.

In  Fig.\ \ref{fig6} we show the 95\% CL exclusion region in the
plane $f_{BB}  \times f_{WW}$ for $M_H=200$ GeV from the study of
reactions (\ref{wwg}) and  (\ref{zzg}). Notice that for these two
reactions the  exclusion region closes the gap at $f_{BB} = -
f_{WW}$ since the anomalous decay widths $H \to W^+W^- (Z^0 Z^0)$
do not vanish along this axis \cite{hagiwara2}. 

We present in Table \ref{tab:f2} the limits on the coefficient
$f/\Lambda^2$ based on a 95\% C.\ L.\ deviation in the total
cross  section for a Higgs mass in the range $170  \leq M_H \leq
350$ GeV. The results coming from the $Z^0 Z^0\gamma$ production
are a little better than the ones obtained from $W^+ W^- \gamma$
production, and they are one order of magnitude better than the
actual limits derived from LEP and Tevatron data analyses.

\section{Discussion}
\label{Conclusions}

So far we have estimated the limits on anomalous dimension--six
Higgs boson interactions that can be derived from the study of
several signatures at LEP and Tevatron colliders. Combined
results from the different reactions were established. We compare
now these results  with existing limits on the coefficients of
other dimension-six operators. 

As discussed in Section \ref{int}, for linearly realized effective
Lagrangians, the modifications introduced in the Higgs  and in
the vector boson sector are related to each other. In
consequence, the bounds on the new Higgs couplings should also
restrict the anomalous gauge--boson self interactions.  Under the
assumption of equal coefficients for all anomalous Higgs
operators, we can relate the common Higgs boson anomalous
coupling $f$ with the conventional parametrization of the vertex
$WWV$ ($V = Z^0$, $\gamma$) \cite{hhpz},
\begin{eqnarray}
\Delta \kappa_\gamma 	&=& \frac{M_W^2}{\Lambda^2}~ f \; ,
\nonumber \\
\Delta \kappa_Z 	&=& \frac{M^2_Z}{2 \Lambda^2}~ (1 - 2 s_W^2)~ f \; ,
\label{trad} \\
\Delta g^Z_1 		&=& \frac{M^2_Z}{2 \Lambda^2}~ f \; .
\nonumber
\end{eqnarray}
A different set of three independent couplings has been also used
by the LEP Collaborations \cite{lepconv}: $\alpha_{B \Phi}$,
$\alpha_{W \Phi}$, and $\alpha_{W}$.  These parameters are
related to the parametrization of Ref.\ \cite{hhpz} through
$\alpha_{B \Phi} \equiv \Delta \kappa_\gamma - \Delta g^Z_1
c_W^2$, $\alpha_{W \Phi} \equiv \Delta g^Z_1 c_W^2$, $\alpha_{W}
\equiv \lambda_\gamma$, or in terms of the anomalous Higgs boson
coupling $f$ by, 
\begin{equation}
\alpha = \alpha_{B \Phi} = \alpha_{W \Phi} = 
\frac{M_W^2}{2 \Lambda^2}~ f =\frac{\Delta\kappa_\gamma}{2}\; .
\end{equation}

The current experimental limit on these couplings from combined
results on double gauge boson production at Tevatron and LEP II
\cite{vancouver} is: 
\begin{equation}
-0.15\;<\;\Delta\kappa_\gamma\;=\;2\alpha \;<\;0.41
\label{wwv}
\end{equation}
at 95 \% CL. This limit is derived under the relations given in
Eq.\ (\ref{trad}) \cite{hisz}. 

\begin{figure}
\protect
\epsfxsize=10cm
\begin{center}
\mbox{\psfig{file=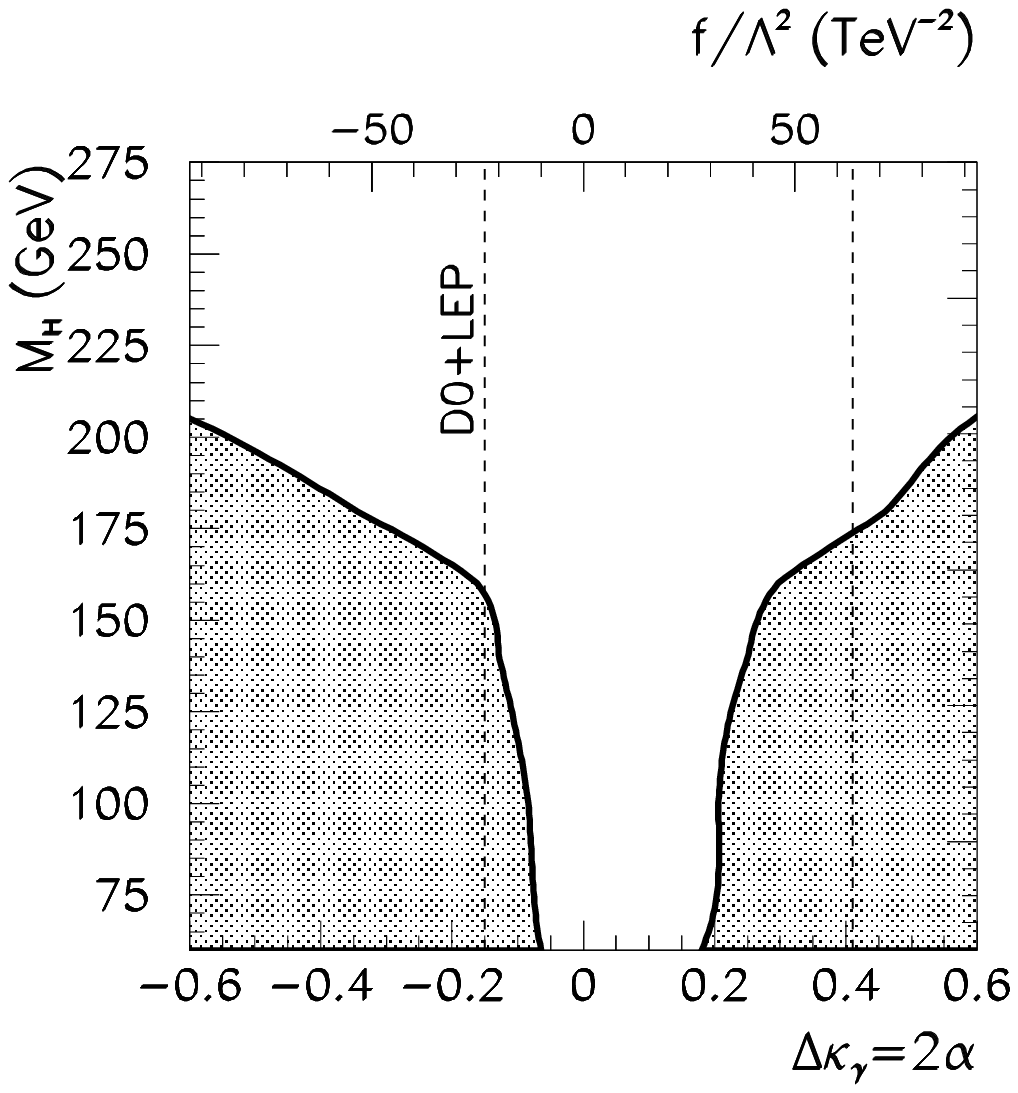,width=0.5\textwidth}}
\end{center} 
\caption{Excluded region in the $f \times M_H$ plane from the
combined analysis of the $\gamma\gamma\gamma$  production at LEP,
$\gamma\gamma\gamma$, $\gamma\gamma + \not \!\! E_T $, and
$\gamma\gamma j j $ production at Tevatron, assuming that all
$f_i$ are equal (see text for details).}
\label{kappa}
\end{figure}

\begin{figure}
\begin{center}
\mbox{\epsfig{file=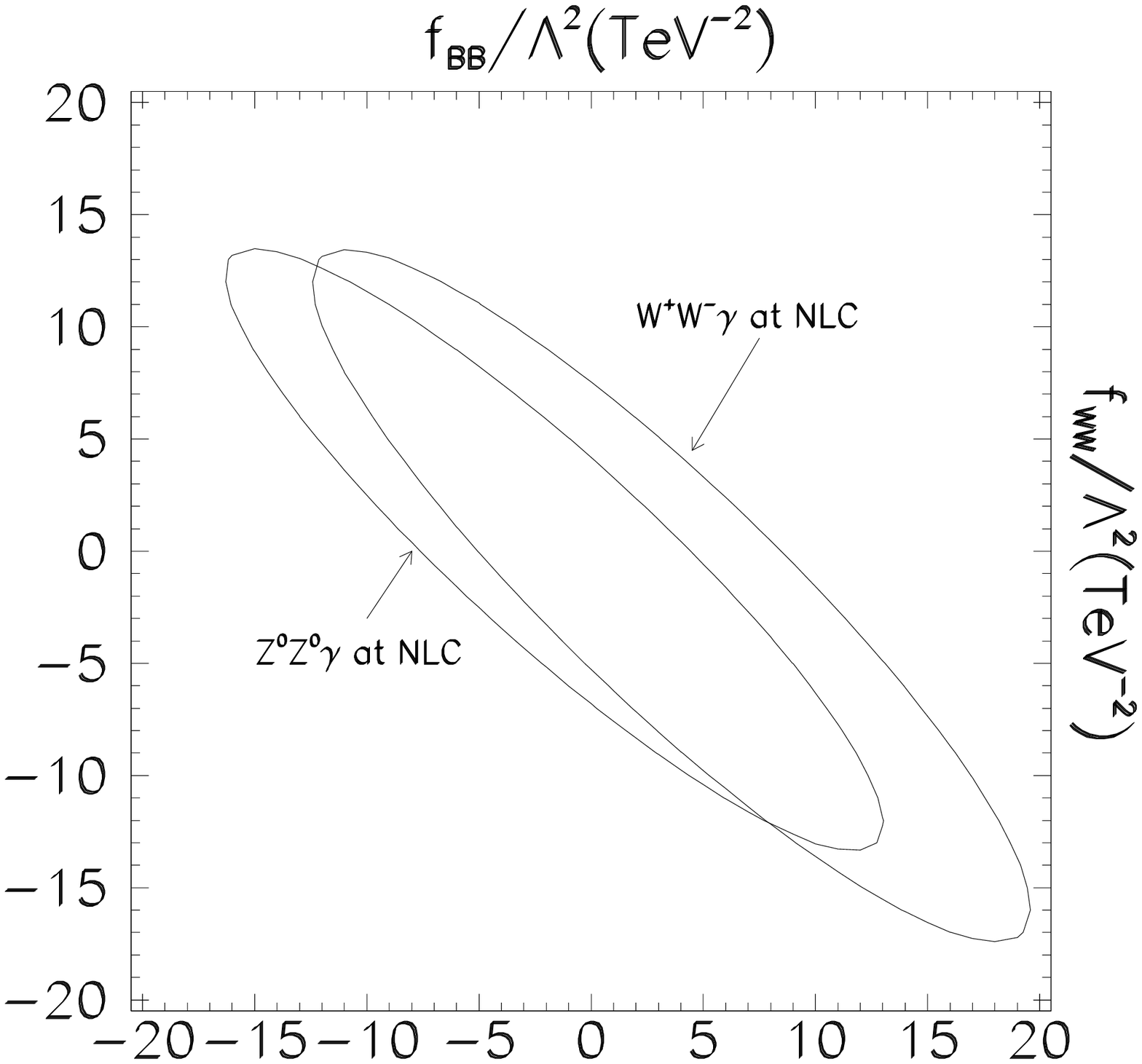,width=0.5\textwidth}}
\end{center}
\caption{Contour plot of $f_{BB} \times f_{WW}$, from $e^+ e^-
\to W^+ W^- \gamma$ (black line) and $e^+ e^- \to Z^0 Z^0 \gamma$ 
(red line) at NLC, for $M_H =200$ GeV with a cut of $p_{T_\gamma} 
> 100$ GeV. The curves show the 95\% CL deviations from the SM 
total cross section.}
\label{fig6}
\end{figure}

In Table \ref{tab:f5}, we present the 95\% CL limit of the
anomalous coupling $\Delta\kappa_\gamma$ using the limits on
$f/\Lambda^2$ obtained  through the analysis of the processes
considered in Section \ref{today}. We also present the expected
bounds  that will be reachable at the upgraded Tevatron and at
the NLC. Our results show that the present combined limit from
the Higgs production analysis obtained in this paper is
comparable with the existing bound from gauge boson production
(\ref{wwv}) for $M_H \leq 170$ GeV, as can be seen in Fig.\
\ref{kappa} (lower scale).

Summarizing, we have estimated the limits on anomalous
dimension--six Higgs boson interactions that can be derived from
the investigation of three photon events at LEP2 and Tevatron and
diphoton plus missing transverse energy events or dijets at
Tevatron. Under the assumption that the coefficients of the four
``blind'' effective operators contributing to Higgs--vector boson
couplings are of the same magnitude, the study can give rise to a
significant indirect limit on anomalous $WWV$ couplings.  We have
also studied the expected improvement on the sensitivity  to
Higgs anomalous couplings at the  Fermilab Tevatron upgrades and
at the Next Linear Collider.


\acknowledgments
M.C. G-G is very grateful to the Instituto de F\'{\i}sica
Te\'orica of Universidade Estadual Paulista for their kind
hospitality.  We would like to thank Alexander Belyaev for very
useful discussions. This work was supported by Conselho Nacional
de Desenvolvimento Cient\'{\i}fico e Tecnol\'ogico (CNPq), by
Funda\c{c}\~ao de Amparo \`a Pesquisa do Estado de S\~ao Paulo
(FAPESP), and by Programa de Apoio a N\'ucleos de Excel\^encia
(PRONEX).


\newpage
\widetext
\newpage 
~
\newpage
\widetext
\widetext

\newpage
~

\newpage

\widetext
\begin{table}
\begin{tabular}{||c||c|c|c|c||}
$M_H$(GeV) & \multicolumn{4}{c||}{$f/\Lambda^2$(TeV$^{-2}$)} \\
\hline 
\hline
  & $e^+ e^- \to \gamma \gamma \gamma$ at LEP & $p \bar{p} \to \gamma \gamma 
  \gamma$ at CDF & $p \bar{p} \to \gamma \gamma +  \not \!\! E_T $ at D\O ~&
  $p \bar{p} \to \gamma \gamma j j$ at D\O ~\\
\hline 
\hline
100 & ( $-$64 , 57 ) & ( $-$62 , 65 ) & ( $-$28 , 57 ) & ( $-$16 , 42 )\\
\hline
120 & ( $-$82 , 70 ) & ( $-$76 , 77 ) & ( $-$37 , 62 ) & ( $-$19 , 46 )\\
\hline
140 & ( $-$192 , 175 ) & ( $-$92 , 93 ) & ( $-$48 , 72 ) & ( $-$26 , 49 )\\
\hline
160 &    ---    & ( $-$113 , 115 ) & ( $-$62 , 84 ) & ( $-$33 , 56 )\\
\hline
180 &    ---    &    ---    & ( $-$103 , 123 ) & ( $-$63 , 81 )\\
\hline
200 &    ---    &    ---    & ( $-$160 , 164 ) & ( $-$96 , 99 )\\
\hline
220 &    ---    &    ---    &    ---    & ( $-$126 , 120 )
\end{tabular}
\medskip
\caption{95\% CL allowed range for $f/\Lambda^2$, 
from  $\gamma\gamma\gamma$ production at LEP OPAL data and 
Tevatron CDF data analysis,  from  $\gamma\gamma + \not \!\! E_T $
Tevatron D\O ~data analysis, and from  $\gamma\gamma j j $
Tevatron D\O ~data analysis assuming all $f_i$ to be equal.
We denote by --- limits worse than $|f|=200$ TeV$^-2$.}
\label{tab:f1}
\end{table}

\widetext
\begin{table}
\begin{tabular}{||c||c|c|c|c||}
$M_H$(GeV) & \multicolumn{4}{c||}{$f/\Lambda^2$(TeV$^{-2}$)} \\
\hline 
\hline
  & $p \bar{p} \to \gamma \gamma \gamma$  & $p \bar{p} 
  \to \gamma \gamma +  \not \!\! E_T $  &
  $p \bar{p} \to \gamma \gamma j j$  & Combined\\
\hline 
\hline
100 & ( $-$24 , 24 ) [ $-$13 , 15 ] 
& ( $-$16 , 36 ) [ $-$9.4 , 26 ] 
& ( $-$9.2 , 22 ) [ $-$3.3 , 5.6 ] 
& ( $-$7.6 , 19 )[ $-$3 , 5.6 ]
\\
\hline
120 & ( $-$26 , 26 ) [$-$14 , 14 ] 
& ( $-$20 , 39 )  [ $-$15 , 27 ]
& ( $-$8.6 , 21 ) [ $-$3.4 , 5.9 ] 
& ( $-$7.4 , 18 )[$-$3.3 , 5.9]
\\
\hline
140 & ( $-$30 , 31 )[ $-$15 , 16] 
& ( $-$25 , 44 ) [ $-$14 , 30]
& ( $-$10 , 23 )  [ $-$4.5 , 8.9]
& ( $-$9.1 , 20 )[ $-$4.0 , 8.7]
\\
\hline
160 & ( $-$36 , 38 ) [$-$17 , 19] 
& ( $-$29 , 50 )  [$-$14 , 33]
& ( $-$11 , 24 )  [$-$6.0 , 14]
& ( $-$9.9 ,22 )  [$-$5.1 , 13]
\\
\hline
180 &    ---   
& ( $-$63 , 72 )   [ $-$46 , 53 ]
& ( $-$26 , 34 )   [ $-$16 , 24 ]
& ( $-$24 , 33 )   [ $-$16 , 24 ]  \\
\hline
200 &    ---     
& ( $-$87 , 90 ) [$-$50 , 53]
& ( $-$33 , 40 ) [ $-$17 , 23]
& ( $-$32 , 39 ) [ $-$17 , 23 ] \\
\hline
220 &    ---    
&    ---     
& ( $-$42 , 45 ) [$-$19 , 26]
& ( $-$42 , 45 ) [$-$19 , 26 ] 
\end{tabular}
\medskip
\caption{95\% CL allowed range for $f/\Lambda^2$, 
from  $\gamma\gamma\gamma$, $\gamma\gamma + \not \!\! E_T $, 
$\gamma\gamma j j $ production at Tevatron Run II [TeV33] assuming  
all $f_i$ to be equal. 
We denote by --- limits worse than $|f|=100$ TeV$^-2$.}
\label{tab:run2}
\end{table}
\widetext
\begin{table}
\begin{tabular}{||c||c|c||}
$M_H$(GeV) & \multicolumn{2}{c||}{$f/\Lambda^2$(TeV$^{-2}$)} \\
\hline 
\hline
  & {$e^+ e^- \to W^+ W^- \gamma$ at NLC} & 
  {$e^+ e^- \to Z^0 Z^0 \gamma$ at NLC} \\
\hline 
\hline
170 & ( $-$2.3 , 3.7 ) &    ---    \\
\hline
200 & ( $-$3.2 , 4.0 ) & ( $-$2.6 ,3.9  ) \\
\hline
250 & ( $-$4.3 , 4.8 ) & ( $-$3.2 , 4.3 ) \\
\hline
300 & ( $-$6.3 , 6.3 ) & ( $-$4.7 , 5.2 ) \\
\hline
350 & ( $-$12 , 9.5 ) & ( $-$7.1 , 8.3 )
\end{tabular}
\medskip
\caption{95\% CL allowed range for $f/\Lambda^2$, 
from $W^+ W^- \gamma$ and 
$Z^0 Z^0 \gamma$ production at NLC, assuming all $f_i$ 
to be equal.}
\label{tab:f2}
\end{table}

\widetext
\begin{table}
\begin{tabular}{||c||c|c||}
Process & $M_H$ (GeV) & $ \Delta\kappa_\gamma=2\,\alpha= 2\,\alpha_{B \Phi} = 
2,\alpha_{W \Phi}$ \\ 
\hline 
\hline
Combined Tevatron RunI + LEPII & 100 &( $-$0.084 , 0.204 ) \\
\hline
Combined Tevatron RunII & 100 & ( $-$0.048 , 0.122 ) \\
\hline
Combined Tevatron TeV33 & 100 & ( $-$0.020 , 0.036 ) \\
\hline
$e^+ e^- \to W^+ W^- \gamma$ at NLC & 200 & ( $-$0.020 , 0.026 ) \\
\hline
$e^+ e^- \to Z^0 Z^0 \gamma$ at NLC & 200 & ( $-$0.016 , 0.024 ) \\
\end{tabular}
\medskip
\caption{95\% CL allowed range for the anomalous triple gauge boson 
couplings derived from the limits obtained for the anomalous Higgs boson 
coupling $f$. }
\label{tab:f5}
\end{table}

\end{document}